\def\be {\begin{equation}}
\def\ee {\end{equation}}

\def\R{{\bf R}}
\def\haux {{\cal H}_{aux}}
\def\hphys {{\cal H}_{phys}} 
\def\sign {{\rm sign}}

\documentstyle[preprint,eqsecnum,aps]{revtex}
\begin{document}
\tighten
\preprint{UCSBTH-96-01, gr-qc/9602019}
\title{Path Integrals and Instantons in Quantum Gravity:
Minisuperspace Models}
\author{Donald Marolf}
\address{Physics Department, The University of California,
Santa Barbara, California 93106} \date{February, 1996}
\maketitle

\begin{abstract}
While there does not at this time exist a complete canonical theory of
full 3+1 quantum gravity, there does appear to be a satisfactory canonical
quantization of minisuperspace models.  The method requires no `choice of time
variable' and preserves the systems' explicit reparametrization invariance.
In the following study, this canonical formalism is used to derive a path
integral for quantum minisuperspace models.  As it comes from a well-defined
canonical starting point, the measure and contours of integration are
specified by this construction.  The properties of the resulting path
integral are analyzed, both exactly and in the semiclassical limit. 
Particular attention is paid to the role of the (unbounded) Euclidean
action and Euclidean instantons are argued to contribute as $e^{-|S_E|/\hbar}$.
\end{abstract}

\vfil
\eject
\section{Introduction}
\label{intro}

Although a complete theory of 3+1 quantum gravity is not yet available,
interesting arguments for the pair production of black holes
\cite{pair1,pair2,pair3}, 
the instability of certain vacuum states \cite{vac}, and
even for a preferred `ground state' of the theory \cite{nb1,nb2}
can be made by analogy with field theories that are better
understood.  A common tool in such arguments is 
a path integral representation of a 
quantum gravity transition amplitude and/or a related semi-classical
approximation.  Our goal here will be to 
investigate this representation and the
$\hbar \rightarrow 0$ limit by using an analogy with a different
class of systems:  
the `finite dimensional
reparametrization invariant models.'

While they are
explicitly defined to have a finite number of degrees of freedom,
such models possess many of the properties which distinguish Einstein-Hilbert 
gravity from the more common quantum mechanical systems or field
theories.  They are invariant under `time reparametrizations,'
a gauge transformation that mixes coordinates and momenta, and as a
result they often possess constraints that are second order in
momenta.  Such `Hamiltonian constraints'
are reminiscent of the `Wheeler-DeWitt equation'
\cite{tril} of quantum gravity.  In addition, they
share with Einstein-Hilbert 
gravity the property that neither the Hamiltonian
function nor the Euclidean action are bounded below.

Since a bounded Euclidean action is required for 
common arguments involving analytic continuation to
Euclidean time, this property has raised 
concern about how a path integral for gravity might be defined 
and analyzed\cite{crot,steve}. One proposal 
\cite{crot,ecr1,ecr2} is to `rotate the contour of integration'
in the path integral until the Euclidean action becomes positive 
definite.  We will study this question below and compare our results to
what would follow from contour rotation.

An important and related point concerns the semiclassical approximation.
Recall that the Einstein-Hilbert action takes the form
\be
\label{EH}
S = \int_{\cal M} \sqrt{-g} R + boundary \ terms
\ee
where $g$ is the determinant of some metric and $R$ is its scalar
curvature.  One might therefore expect that quantum gravity
path integrals can be treated as integrals of analytic functions and 
that the contour can be deformed from the original region of integration.
Some arguments for the contribution of Euclidean instantons can be made in this
way, by deforming Lorentzian metrics to Euclidean ones through the
complex plane.  However, once deformations into the complex plane
are allowed, one must in principle include in the semiclassical
analysis {\it any} stationary points which lie within the domain
of analyticity of the integrand.  Because of the branch cut
introduced by the $\sqrt{g}$ factor in \ref{EH}, the metric
$g$ effectively takes values on a Riemann surface and, at least
for appropriate boundary conditions, the stationary
points occur in pairs {\it with opposite signs of the action.}
This point has been made many times \cite{regge,HH,JJ1,JJ3} and a
similar feature arises in our finite-dimensional reparametrization
invariant models.  Which stationary phase points actually contribute
to the result will depend on the particular contour of 
integration.  This is, however, an issue that must be analyzed and one
is lead to wonder what determines this contour so that, for example, 
black hole pair creation calculations predict 
an exponentially suppressed rate for large mass and not one that is
exponentially enhanced.
 
Our goal here is to investigate these issues in the finite dimensional
reparametrization invariant context by deriving our path integral
from a canonical quantum formalism.  The existence
of a canonical formalism which can be applied to a large class of 
models and which does not require `deparametrization' or the
imposition of gauge fixing conditions is a fairly recent development
\cite{lands,qord,BIX,ash,ban} whose implications for path 
integral methods have not yet been explored.  Such is our aim here, 
and we will find that the resulting path integral 
has two interesting properties:  First, it can be written as an 
integral over both `Lorentzian' and `Euclidean' paths in
which the contribution of Euclidean paths is always exponentially
{\it suppressed} for small $\hbar$.  Second, the Euclidean semiclassical
approximation yields only exponentially damped contributions;
in particular, Euclidean instantons always contribute
with the weight $e^{-|S_E|/\hbar}$, where $S_E$ is the Euclidean 
action of the instanton.

We begin by describing the aforementioned canonical formalism
in section \ref{cf}.  This brief review is
intended to provide a working understanding
of the scheme without addressing all of the technical details or
supplying all of the motivations.
A discussion of these points can be found in the
literature \cite{lands,qord,ash,ban}.

After describing the the canonical formalism, we proceed in 
section \ref{pi} to define our transition amplitude and to 
express it as a path integral.  This transition amplitude is then
evaluated in section \ref{exact} for two classes of exactly
solvable models.  In both, it is clear that `Euclidean' semiclassical
contributions are exponentially suppressed for large
$|S_E|$.  Section \ref{conv} rewrites the
path integral in a more strongly convergent form which may be useful 
for numerical investigations.  Section then
\ref{inst} addresses the semi-classical approximation.  We
close with a brief discussion in section \ref{disc}.

\section{The Canonical Formalism}
\label{cf}

We now describe the specific class of models to be addressed and
review the canonical formalism of \cite{lands,qord,ash} on which our
path integral will be based.  The goal of this review is to provide
a working understanding of the scheme and not to describe
the technical subtleties in detail.  For a more complete and rigorous
presentation, see \cite{lands,ash,ban} and for 
a discussion of existence and uniqueness issues, see \cite{ban}.  Essentially
the same construction was introduced in
\cite{hig,htip} in slightly different contexts. 

In this paper, we will study only 
finite dimensional systems possessing a single
constraint proportional to the system's Hamiltonian. 
We take the system to be 
expressed in canonical language in terms of a phase space
$\Gamma$, which we shall assume to be $T^*\R^n$ and to be 
equipped with coordinates $x^i, p_i, i \in {0 .... n-1}$.  
Finally, we assume that the
constraint is of the form
\be
\label{const}
h = g^{ij}(x)p_ip_j + V(x).
\ee
As is well known, many such systems arise as
`mini-superspace models' of gravitating systems \cite{misner,MTW,Jan,RS,atu} 
and so form a set of some interest.
Note, however, that the method applies to more general
models \cite{lands,qord,ash,ban} and has been used to construct
states of linearized gravity \cite{hig} as well as a Hilbert 
space of `diffeomorphism-invariant' states
\cite{ash} in the loop representation approach to quantum gravity.

The quantization scheme to be followed here is known as
the `refined algebraic method' (which is closely related to the
`Rieffel induction method' of \cite{lands})
and may be thought of as an elaboration of
the Dirac approach \cite{yesh}, in which the constraints are 
required to annihilate the so-called `physical states.'  Specifically,
the refined algebraic approach asks that we first quantize the system
while completely ignoring the constraint.  This provides an 
`auxiliary' Hilbert space $\haux$ in which to work.  This space
is called auxiliary because it contains much more than the 
`physical states' that satisfy the constraints.  In our case, we will
take this space to be $\haux =  L^2(\R^n)$ with operators
$X^i$ (coordinates) and $P_i$ (momenta) 
acting in the usual way.  For simplicity in our expressions, we 
define our $L^2$ space using the measure $d^nx$.  In this paper we
follow the convention of \cite{qord} in denoting classical phase space
functions by lower case letters while denoting
quantum operators 
by capital letters.  We use units in which $\hbar =1$, except in
section \ref{inst} which explicitly investigates the $\hbar 
\rightarrow 0$ limit.

The next step in the procedure is to `quantize' the constraint $h=0$.
For our purposes, this simply means that we choose some
self adjoint operator $H$ on $\haux$ which has the function $h$ as
its classical limit.  The usual ambiguities are present at this
level and we make no attempt to give a unique prescription.
In fact, a somewhat greater ambiguity is present
here than in quantizing the Hamiltonian of a non-relativistic 
system.  The point is that, classically, the constraint $h = 0$ 
is equivalent to any constraint of the form $f(x)h = 0$, 
although such a rescaling 
{\it can} affect our quantum prescription when $f(x)$ is not a 
constant\footnote{In particular, the proposals of \cite{Hal,conf1,conf2} are not
compatible with our choice of inner product.}.
Our viewpoint here is that this is just one more of the many ambiguities
that arise when a classical system is quantized.  

Now, if the spectrum of $H$ were entirely discrete, the
implementation of the Dirac prescription would be straightforward.
Those eigenstates of $H$ with eigenvalue zero would become the 
physical states of our theory and the `physical Hilbert space'
could simply be the $H=0$ eigenspace of $\haux$.  However, in 
typical cases $H$ will also have a continuous
spectrum at zero eigenvalue, for which the corresponding eigenstates
will not be normalizable in the auxiliary Hilbert space but will
instead be `generalized eigenstates' of $H$, a kind of distribution.

The strength of the refined algebraic quantization procedure is its
ability to form a physical Hilbert space from such generalized
states using an inner product that is, in a certain sense, induced
from $\haux$.  This has the advantage that any
sufficiently `nice' operator $A$ on $\haux$ which commutes with $H$ 
induces a densely defined operator $A_{phys}$ on the physical Hilbert
space $\hphys$.  In addition, the map $A \mapsto A_{phys}$ is a
`*-algebra homomorphism,' preserving multiplication, addition, and
Hermitian conjugation of the operators.  Since $\haux$ is a 
quantum version of the phase space $\Gamma$, it is through
this induction process and the auxiliary Hilbert space $\haux$
that the *-algebraic properties of the observables on
$\hphys$ are connected to the reality properties of the classical
phase space functions. 
This is in fact the important point, as the most `physical'
requirement of an inner product is that it give the
proper adjointness relations to the quantum operators \cite{AAbook,AR}.
This construction is described below.
The reader is encouraged to consult
\cite{qord,ash,ban} for further details.

We shall in fact assume the spectrum of $H$ to be {\it entirely}
continuous at $H=0$.  That this case is in some sense sufficient
follows from the result \cite{ash,ban} that the continuous
and discrete eigenstates of $H$ induce sectors of the
physical Hilbert space which are {\it superselected} relative to each other.
The presence of discrete eigenstates would, however, affect
the formulation of the path integral in ways that we would prefer
to ignore.  We therefore content ourselves with the observation
that many minisuperspace models can be formulated with a
constraint having only continuous spectrum at $H=0$ and restrict
to this case; for
details see \cite{qord} and in particular \cite{BIX} 
for the case of the Bianchi IX model.

In this situation and under a certain technical assumption 
concerning the operator $H$, the physical Hilbert space is straightforward
to construct.  What we would really like is to `project'
$\haux$ onto the (generalized) states which are zero-eigenvalue eigenvectors of
$H$.  Of course, since none of these states are normalizable, this
will not be a projection in the technical sense.  Instead, it will
correspond to an object which we will call $\delta(H)$, a Dirac
delta `function.'  Given the above mentioned assumption on
$H$ (see \cite{ban}), the object $\delta(H)$ can be shown to exist
and to be uniquely defined.  Technically speaking however, it 
exists not as an operator in the Hilbert space $\haux$, but as
a map from a dense subspace ${\cal S}$ of $\haux$ to the (for our
purposes, topological) dual
${\cal S}'$ of ${\cal S}$.  The space ${\cal S}$
may typically be thought of as a Schwarz space; that is, as the space of
smooth rapidly decreasing functions on the configuration space.
In this case, ${\cal S}'$ is the usual space of tempered distributions.
Not surprisingly, this is reminiscent of 
the study of generalized eigenfunctions through Gel'fand's 
spectral theory \cite{Gelfand} and ${\cal S} \subset \haux
\subset {\cal S}'$ forms a rigged Hilbert triple.

The key point is then as follows.  While generalized eigenstates
of $H$ do not lie in $\haux$, they 
can be related to normalizable
states through the action of the `operator' $\delta(H)$.  That is, 
generalized eigenstates $|\psi_{phys}\rangle$ of $H$ with 
eigenvalue $0$, can always be expressed in the form $\delta(H)|\psi_0
\rangle$, where $|\psi_0\rangle$ is a normalizable state in
${\cal S} \subset \haux$.  This choice of $|\psi_0\rangle$ is of course not
unique and, in fact, we associate with the physical state $|\psi_{phys}
\rangle$ the entire {\it equivalence class} of normalizable
states $|\psi\rangle \in {\cal S}$ satisfying
\be
\delta (H) |\psi \rangle = |\psi_{phys} \rangle.
\ee
Each equivalence classes of normalizable states will form a {\it
single} state of the physical Hilbert space.

All that is left now is to `induce' the physical inner product from
the auxiliary Hilbert space.  Naively, the 
inner product of two physical states $|\phi_{phys}\rangle$ and
$|\psi_{phys} \rangle$ may be written $\langle \phi | \delta(H)
\delta (H) | \psi \rangle$, where $|\phi \rangle$ and $|\psi \rangle$
are normalizable states in the appropriate equivalence classes.
This inner product is clearly divergent, as it contains
$[\delta(H)]^2$.  The resolution is simply to `renormalize' this
inner product by defining the {\it physical} inner product to be
\be
\label{pip}
\langle \phi_{phys}| \psi_{phys} \rangle_{phys} =  \langle
\phi | \delta(H) | \psi \rangle_{aux},
\ee
where the subscripts $phys$ and $aux$ on the brackets indicate the
two different inner products.
Note that \ref{pip} does not depend on
which particular states
$|\phi \rangle$, $|\psi \rangle \in {\cal S}$ were chosen to represent
the physical
states $|\phi_{phys}\rangle$ and $|\psi_{phys}\rangle$.  
This construction parallels the 
case of purely discrete spectrum as, if $P_H$ were a projection
onto normalizable zero-eigenvalue eigenstates of $H$, we would have
$[P_H]^2 = P_H$.  Although $\delta(H)$ is not strictly speaking
an operator, taking $|\phi \rangle $ and $|\psi \rangle$
to lie in ${\cal S}$ makes the above inner product well defined, as
well as Hermitian.  In the case of the free relativistic
particle, this positive definite 
inner product corresponds to the Klein-Gordon
inner product on the positive frequency states, but corresponds
to {\it minus} the Klein-Gordon inner product on the negative frequency 
states.  The positive and negative frequency subspaces are orthogonal
as usual.  A similar representation of the inner product holds in certain 
other cases \cite{jim}.  

Perhaps the most important feature of this approach is that
it  first defines the algebra of quantum observables (without requiring them to
be found explicitly) and then provides a *-representation of this
algebra on the physical Hilbert space; that 
is, a representation on $\hphys$ in which the proper Hermitian
conjugation relations hold.  
{}From the algebraic point of view, this is the fundamental goal of any
quantization scheme, and it is this *-representation that
determines all physical predictions.  

The representation is defined as
follows.  The observable algebra is defined by the *-algebra of
observables that commute with the constraint $H$ and act `nicely' on
the dense set ${\cal S}$ (see \cite{ash,ban} for details).  
These are the analogues of the (smooth)
gauge invariants of classical physics.  Each such operator
$A$ then
{\it induces} an operator $A_{phys}$ on $\hphys$ through
\be
A_{phys} |\psi_{phys} \rangle \equiv \delta(H) A |\psi \rangle,
\ee
where again $|\psi\rangle$ is any state for which $|\psi_{phys} \rangle
= \delta(H) |\psi \rangle$.
The operators
$A_{phys}$ then satisfy the same algebraic and Hermitian conjugation
relations as the observables on $\haux$, forming the desired
*-representation.  The use of such operators and the
physical inner product  \ref{pip}
has been shown to give physically reasonable results in interesting 
special cases \cite{qord,BIX} and in the semiclassical
limit \cite{bdt}.

\section{A Path Integral for the Inner Product}
\label{pi}

Having described how our models will be quantized,
we now wish to derive a path integral formalism
for these systems.  To do so, we must first answer the 
question ``Just what quantity
should we derive a path integral for?"  Path integrals are often used
to represent the `transition amplitudes' that encode the time
evolution of quantum systems.  However, for the cases we consider, 
the Hamiltonian explicitly vanishes on the physical Hilbert
space.  Thus, the operator $e^{-iHt}$ is just the identity.

Nevertheless, we know that the physical states {\it do} contain
information that we may call dynamical (see, for example, 
\cite{bryce,Carlo,Carlo2,Carlo3,Carlo4} 
for general comments or  \cite{qord,bdt} for a 
discussion in the context of this particular approach).  Thus, there
should be some mathematical object which, more or less, encodes
our idea of a `transition amplitude.'  

Apparently nontrivial path integral expressions for `transition
amplitudes'
have in fact been studied by a number of authors (e.g., \cite{JJ1,JJ3,HLH}).
In the minisuperspace context, all of these are transition amplitudes
between two {\it configurations} $x$ and $x'$ (which are the 
analogues of the three-geometries of $3+1$ gravity), or perhaps
their conjugate momenta.  
As a result, we will seek our transition amplitude in the 
auxiliary Hilbert space $\haux$.

We take our `transition amplitudes' 
to be just the matrix
elements of the `operator' $\delta(H)$ in $\haux$.
That is, we will compute $\langle x|\delta(H)|x'\rangle$
where $|x\rangle$ and $|x'\rangle$ are 
generalized eigenstates of the coordinate
operators $X^i$.  Our reasons for this are two-fold.
First, while expressed in terms of the auxiliary space, 
such matrix elements contain all of the information about $\hphys$ 
as they define the physical inner product.  
Second, when one of the coordinates (say $x^0$) is considered to 
represent a `clock' and when this clock behaves semi-classically 
\cite{bdt} this object does in a certain
sense describe the amplitude for the `evolution' of the `state' 
$\underline{x}$ at `time' $x^0$ to the `state' $\underline{x'}$ at
`time' $x^0{}'$.  Here $\underline{x}$ represents the coordinates on
a slice through the configuration space of constant $x^0$.  

It is now straightforward to represent this object as a path integral.
To do so, consider the path integral expression for the operator
$e^{-iNH}$ on $\haux$, which we expect to exist and which can 
be derived in the usual way by skeletonization (see, for 
example \cite{HT}) of paths between $x$ at `time' zero 
and $x'$ at `time' $N$.  Note that by `time' we mean an additional
parameter that we now introduce; {\it not} one of the coordinates
$x^i$.  We then integrate
$N$ from $-\infty$ to $\infty$ to turn $e^{-iHN}$ into $\delta(H)$.

The resulting path integral is then
\be
\label{mepi}
\langle x | \delta(H) | x \rangle = {1 \over {2 \pi}}
\int_{-\infty}^{\infty} dN \int Dx Dp 
\exp \left( i \int_0^N dt \left[ p\dot{x} - h(x(t),p(t)) \right] \right)
\ee
where $\int_{-\infty}^{\infty} dN$ denotes an  integral over the {\it
single} variable $N$ while $Dx Dp$ denotes the Liouville measure on
the canonical path space \cite{HT}.  Since the configuration variables are
specified at the endpoints, there is `one less' set of integrations
over the coordinates than over the momenta.  The notation $h(x(t),p(t))$
stands for the `symbol' ${{\langle p = p(t) | H | x = x(t) \rangle_{aux}}
\over {\langle p | x \rangle_{aux}}}$
of the operator $H$, where $|p\rangle$ is an eigenstate of the
momenta $P_i$.

In the usual way, gauge fixing machinery and redundant integrations
can now be introduced to write \ref{mepi} in a form where its
independence of `gauge' is more explicit.  Halliwell \cite{Hal}
has performed this analysis in the reverse direction by starting
with the Fadeev-Popov form and introducing the gauge
fixing condition $\dot{N} = 0$.  Since he arrives at just
our result \ref{mepi}, his work allows us to express the physical
inner product in the form
\be
\label{mepiN}
\langle x | \delta(H) | x \rangle = {1 \over {2 \pi}}
\int DN Dx Dp  \ \delta(G) \Delta(G)
\exp \left( i \int_0^1 dt  \left[ p\dot{x} - N(t) h(x(t),p(t)) \right]
\right)
\ee
where $G$ is now any appropriate gauge fixing function, $\Delta(G)$ is the
associated Fadeev-Popov determinant, and the sum is over all
paths $(x(t),p(t),N(t))$ in which $x^i(t),p_i(t),N(t)$ are
allowed to range over the entire real line.
This is the path integral that we shall explore in section 
\ref{gen}.  For interested readers, the convergence properties of
\ref{mepi} and \ref{mepiN} are discussed in detail in appendix 
\ref{cds}.

\section{Exactly Solvable Models}
\label{exact}

Having derived a path integral for 
$\langle x | \delta(H) | x' \rangle$, the `physical inner product,'
it is of interest to see what form this distribution takes in simple
cases where an exact analytic expression can be obtained.  As usual, the
cases that we will study are the `purely quadratic ones;' 
the perturbed Bianchi I model (or free relativistic particle) and the case of 
coupled Harmonic oscillators.

\subsection{Perturbed Diagonal Bianchi I}
\label{frp}

The Bianchi I model is a minisuperspace 
describing spatially homogeneous spacetimes of the form ${\cal M} =
T^3 \times \R$ having a 
foliation by three-tori with flat Riemannian metrics (so that the
tori form spacelike hypersurfaces of ${\cal M}$).  In the
diagonal version of this model the metric is such that, at each
point $x \in {\cal M}$, three mutually orthogonal closed
geodesics intersect at
$x$ and each encircle an arm of the torus once.  This system
may be formulated on the
configuration space ${\cal Q}=\R^3$ with a constraint 
of the form
\be
h_{BI} = - p_0^2 + p_1^2 + p^2_2. 
\ee
In this case, 
the coordinate $x^0$ describes the volume of the three-torus 
while the coordinates $x^1$ and $x^2$ describe the
`anisotropies,' the ratios of the lengths of minimal curves encircling 
the torus in different directions.  We will consider
a `perturbed' and slightly
more general model on
${\cal Q} = \R^n$ for which the quantum constraint is
\be
H = {1 \over 2} (- P_0^2 + \sum_{i=1}^{n-1} P_i P_i + m^2)
\ee
for $m^2 > 0$.  Note that without the perturbation ($m^2$), the
action $S$ would vanish on every classical solution and we would be
unable to consider the semiclassical limit $|S| \gg 1$.
The constant factor of ${1 \over 2}$ affects
none of the results, but conforms to the usual convention for the
normalization of kinetic terms.

Such a system looks exactly like the
free relativistic particle.  However, we will intentionally
avoid referring to this system by that name, as we feel that the
{\it physics} of the two situations is quite distinct.  This follows
from the fact that the metric which defines the constraint's `kinetic
term' has a different interpretation in each of the two cases.
A free relativistic particle with $p_0 <0$ is usually
interpreted as `traveling backwards in time,' a process that we
suppose to be physically disallowed.  This leads to the usual
preference for positive frequency states over negative frequency
states.  However, in the Bianchi I
model, a negative $p_0$ means only that at the $x^0$ of the tori
is decreasing with proper time -- that is, that the universe is {\it
collapsing}.  This is not only a physically interesting process but a
process which classically {\it must} occur in some minisuperspace
models, such as Bianchi IX \cite{Wald}.  Thus, we are pleased to include
the negative frequency states in our model.

We now proceed to compute the integral
\be
\label{frpint}
\langle x | \delta(H) |x ' \rangle = {1\over {2 \pi}} 
\int_{-\infty}^{\infty} dN \langle x | e^{-iHN} | x' \rangle =  {1 
\over \pi} \Re ( \int_0^{\infty}  \langle x | e^{-iHN} | x' \rangle )
\ee 
where $\Re$ denotes the real part.
The operator $e^{-iHN}$ is just $e^{-im^2N/2}$ times a product 
of propagators for free nonrelativistic particles.  As a result, its
matrix elements are readily seen to be
\be
\langle x | e^{-iHN} | x' \rangle = e^{{{i\pi} \over 2} (1 - {n \over
2})} (2 \pi N)^{-n/2} \exp{\left[ - {{im^2} \over 2}\left( N 
- {{(x - x')^2} \over {m^2 N}} \right) \right] }
\ee
for $N >0$.
We can evaluate \ref{frpint} using 3.471 of \cite{GR} to
yield, for $(x -x')^2 >0$,
\be
\label{spacelike}
\langle x | \delta(H) |x' \rangle = {2 \over {\pi (2 \pi)^{n/2}}}
\left[{{\sqrt{(x-x')^2}} \over m} \right]^{(1-n/2)} K_{(n/2)-1} \left( m
\sqrt {(x-x')^2} \right)
\ee
where $K_{(n/2)-1}$ is the modified Bessel function (of the second kind)
of order $(n/2)-1$.  Similarly, for $(x-x)^2 < 0$, we find
\begin{eqnarray}
\label{timelike}
\langle x | \delta(H) |x' \rangle &=& {1 \over {(2 \pi)^{n/2}}}
\left[{{\sqrt{-(x-x')^2}} \over m} \right]^{(1-n/2)}
\Bigg\{
\cos{\left[ \pi \left( {{n-1} \over 2} \right) \right]}  J_{(n/2)-1} \left( m
\sqrt {- (x-x')^2} \right) \cr &-& 
\sin{\left[ \pi \left( {{n-1} \over 2} \right) \right]}  N_{(n/2)-1} \left( m
\sqrt {- (x-x')^2} \right) \Bigg\}.  
\end{eqnarray}
An intuitive feel for just why the answer is of this form
may be obtained by noting that it describes a certain
correlation function in free scalar field theory, or alternately by
performing this integral in the
semiclassical approximation.  For the interested reader, the
semiclassical
analysis is carried out in appendix \ref{sc}. 

Note that, for $-(x-x')^2 \gg m^2$, the matrix elements are
roughly $\sin(m \sqrt{-(x-x')^2})$ or  
$\cos(m \sqrt{-(x-x')^2})$ depending on the number of degrees
of freedom.  That is, they
include equal contributions from what might be called the `positive
and negative frequency parts.'  Nevertheless, when $(x-x')^2 \gg m^2$,
the matrix elements contain only the
{\it decreasing} exponential $\exp(-m \sqrt{(x-x')^2})$.  
This occurs
even though the Euclidean action 
is unbounded below.   Similar results hold for the case 
$m^2 = -k^2 <0$ for which the resulting physical
inner product is obtained by replacing $m$ with $k$ and 
$(x-x')^2$ with $-(x-x')^2$ in \ref{spacelike} and
\ref{timelike}.  As might
be expected, this is related to which points of stationary phase
contribute to $\langle x | \delta(x) | x' \rangle$
in the semiclassical approximation.  We will return to this
point in section \ref{gen} and appendix \ref{sc}.

It is interesting to compare the results \ref{spacelike} and 
\ref{timelike} to what one might obtain by a `contour rotation
prescription' of the sort suggested in \cite{crot}. 
The idea
is to rotate the contour of the conformal mode [represented
here by $x^0$] to imaginary values [$x^0 \rightarrow i x^0]$
so that the Euclidean action becomes
bounded below.  We can certainly perform the integral \ref{frpint}
taking $-p_0^2$ to be replaced by $p_0^2$ in the Hamiltonian $h$.
The result is just \ref{spacelike} above with $(x-x')^2$ replaced
by $(x -x')_E^2 \equiv \sum_{i=0}^n (x^i)^2$, since $(x-x')_E^2 \ge 0$.
The task is then to analytically continue back;
$x^0 \rightarrow -i x^0$.
For $(x-x')^2 > 0$, this clearly
yields \ref{spacelike}; the correct answer from our point of view.
However, for $(x-x')^2 < 0$,
the analytic continuation is ambiguous.  Because of the 
branch cut that defines the square root, the answer will depend on which
path is followed through the complex plane.  As can be seen from
\ref{timelike}, what we would call the `right' answer results from
combining these two contributions with equal weights
and with a phase that depends on $n$.

\subsection{Coupled Harmonic Oscillators}
\label{cho}

In order to more thoroughly explore the properties of
exactly solvable models, we now study the case of 
coupled harmonic
oscillators.  However, to fit with the general
character of minisuperspace models, we will take one of the 
oscillators to have negative energy.  That is, we will
again take our system to be defined on the phase space 
$\Gamma = T^*\R^n$, with the constraint 
\be
\label{hoham}
0 = h = {1 \over 2} [-p_0^2 - 
(x^0)^2 \omega^2 + \sum_{i=1}^{n-1} (p_i^2 +
(x^i)^2 \omega^2)] \equiv {1 \over 2} [p^2 + \omega^2 x^2 ]
\ee
for $\omega^2 > 0$.
We could, in principle, also study
the case where one or more of the Harmonic oscillators is inverted
(with Hamiltonian ${1 \over 2} (p^2 - \omega^2 x^2)$), but we will not do 
so here.
For simplicity, we assume $n$ to be even.
Models of the form \ref{hoham} do arise as minisuperspace models of
gravity interacting with scalar fields \cite{homod} and many of their
aspects have been studied in the literature \cite{homod,tate,jorma}.

Strictly speaking, however, this model does {\it not} fall into the
class allowed by section \ref{cf}.   Since it is a sum of harmonic
oscillator Hamiltonians,
instead of the 
constraint $H$ having purely continuous
spectrum at zero (when quantized in the usual way on $L^2(\R^n)$), its
spectrum is purely {\it discrete} -- not just at zero, but everywhere.
This creates a number of subtleties, such as the fact that there
are rather few zero eigenvalue eigenstates of the constraint unless 
all of the oscillators have commensurate frequencies.  Furthermore, 
unless an appropriate constant is added to $H$, it will in general
have {\it no} eigenstates with eigenvalue zero.  Thus, a better
way to quantize this model might be to rewrite the constraint
(see, for example, \cite{qord,BIX,bdt}) in such a way that the spectrum is 
purely continuous\footnote{This continuous spectrum form of the 
constraint is in fact in better accord with the global properties of 
the original gravitational model of \cite{homod}.}.  
Unfortunately, this necessarily destroys the 
purely quadratic form of the action and thus the exact solvability of
the model.

Our purpose in studying this model is not to examine the detailed
predictions of the cosmological scenario of \cite{homod},
but rather to investigate the general mathematical properties of 
the expressions \ref{mepi} and \ref{mepiN} for the physical
inner product.
Thus, we will quantize this model using the constraint  operator
\be
\label{qhoham}
H = {1 \over 2} [-P_0^2 - 
(X^0)^2 \omega^2  + \sum_{i=1}^{n-1} (P_i^2 + 
(X^i)^2 \omega^2)] + [1 - {n \over 2}] \omega. 
\ee
The frequencies and constants in \ref{qhoham} are specifically chosen so
that there seem to be a sufficient number of zero-eigenvalue eigenvectors.
We shall not concern ourselves with whether or not such choices
are `natural.'

However, if we are to proceed in this way, expressions \ref{mepi}
and \ref{mepiN} must be modified.  This is because their
aim was to calculate the operator $\delta(H)$ and, now that the
spectrum of $H$ is discrete, this object is highly divergent:
when acting on a normalizable state $|\psi \rangle$ for which
$H |\psi \rangle = 0$, $\delta(H)$ cannot possibly be defined.

As stated in section \ref{cf}, the analogous object for the case
of discrete spectrum is the projection  $P_{H=0}$
onto the zero eigenvalue subspace for the
operator $H$.  An expression for $\langle x | P_{H=0} | x' \rangle$
analogous to \ref{mepi} and \ref{mepiN} can be found by
realizing that the `evolution operator' $e^{-iNH}$ is, in this case, 
periodic in $N$ with period $2 \pi /\omega$.  
This of course is the source of the would-be
divergence in $\delta(H) = {1 \over {2 \pi}}
\int_{-\infty}^{\infty} dN \ e^{-iNH}$, but it also allows
us to express the projection $P_{H=0}$ as the integral over a single
cycle of $N$:
\be
\label{proj}
P_{H=0} = {{\omega} \over {2 \pi}} \int_{- \pi / \omega}^{\pi / \omega} dN \ 
e^{-iNH} = {\omega \over \pi} \int_0^{\pi/\omega} dN \
\left( e^{-iNH} + e^{iNH} \right).  \ee
Similarly, the analogues of \ref{mepi} and \ref{mepiN} are given by
taking $N$ to live on a periodic interval of length $2 \pi/ \omega$.

As in the perturbed Bianchi I model, the operator
$e^{-iHN}$ is a product of evolution operators for nonrelativistic
particles which can be evaluated exactly; perhaps most easily by
using the 
semiclassical approximation (which is exact for quadratic Hamiltonians).
For ${\pi \over \omega} > N > 0$, the result is 
\be
\label{sce}
\langle x | e^{-iHN} |x' \rangle = \left( {{\omega} \over { 2\pi \ \sin \omega
N}} \right)^{n/2} e^{i{\pi \over 2}(1- {n \over 2 })}
e^{iS_{cl}(x,x';N)}
\ee
where
\be
S_{cl}(x,x';N) =  {{\omega} \over 2} {{ (x^2 + x'{}^2) \cos \omega N
- 2 x\cdot x'} \over \sin \omega N}
\ee
is the action of the least action path between $x$ and $x'$ traversed
in the {\it given}
time $N$.  Note that this path is unique for $N \neq 0, \pi/ \omega$.
Here, $x \cdot x'$ denotes $-x^0x^0{}' + \sum_{i=1}^{n-1} x^ix^i{}'$.

All that remains is to evaluate the integral over $N$ in \ref{proj}.
For the case $n=2$, we may refer to \cite{n2j,n2k}.  More generally,
if we define $a =  - \omega (x^2 + x'{}^2)/2$ and $b  = \omega x
\cdot x'$ then 
the result may be expressed as
\be
\label{home}
\langle x | P_{H=0} | x' \rangle = 2 \omega 
  \Re  \left\{ (2 \pi)^{- (n/2)} i e^{i {\pi \over 2} (1 - {n \over
2})} 
\left( i { {\partial} \over
{\partial b}} \right)^{(n/2) -1}  H_0^{(1)} (\sqrt{b^2 - a^2})
\right\},  
\ee
using 6.677 from \cite{GR}.  Here, $H_0^{(1)}$ is the usual
Hankel function of the first kind and the
square root is defined by $\sqrt{r e^{i \theta}} = r^{1/2} e^{i
\theta/2}$ for $0 \le \theta \le \pi$.  Thus, when
$a^2 > b^2$ \ref{home} takes the form
\be
\label{hokf}
\langle x | P_{H=0} | x' \rangle =
{{4 \omega} \over \pi} (2 \pi)^{-n/2} \left( {\partial \over {\partial
b}} \right)^{(n/2)-1} K_0(\sqrt{ a^2 - b^2}).
\ee
The results \ref{home} and \ref{hokf} may in
turn be written as
a sum of Hankel functions of orders $0 < \nu < n/2-1$ by using the usual
Bessel function recurrence relations.

The matrix elements \ref{home} have the same general
structure as those of \ref{spacelike} and \ref{timelike}.
That is, they are a Bessel function of order $n/2-1$ of the square root
of a function of $x$ and $x'$.  Recall that in the case of 
perturbed Bianchi I, the behavior of this Bessel function was determined
by the sign of $(x-x')^2$; that is, by whether the points $x$ and
$x'$ could be connected by a `Lorentzian' classical solution
(with real lapse $N$) or by a `Euclidean solution' with imaginary
lapse.  A similar phenomenon occurs for our coupled oscillator model.
Pairs of configurations $(x,x')$
may be separated into three distinct classes: those for which
$-{a \over b} >1$, those for which $1 > - {a \over b} > -1$, and 
those for which $ -1 > - {1 \over b}$.  Only for the first
category ($-{a \over b} > 1 $) does a `Euclidean' classical
solution (with imaginary lapse) exist.  Such solutions
come in pairs with Euclidean actions
$S_E = \pm \sqrt{a^2 - b^2}$.  Similarly, a `Lorentzian'
solution (with real lapse) exists only for the second case, 
$1 > - {a \over b} > -1$.  These solutions also come in pairs, with
actions $S = \pm \sqrt{b^2 - a^2}$.  For the third
case ($- {a \over b} < -1$), the configurations $x$ and $x'$
may be connected by  {\it complex} classical solutions with
$\Re N = {\pi \over \omega}$ (but no such solutions exist for
the first two classes of pairs).  The
Euclidean action is also real for this last case, and is again
given by $S_E = \pm \sqrt{a^2 - b^2}$.

The semiclassical approximation is valid (in our units) when $|S| =
\sqrt{|b^2 -
a^2|} \gg 1$ so that, once again, when $x$ and $x'$
are connected by a Lorentzian solution,
this approximation contains contributions both of the form $e^{iS}$
and $e^{-iS}$ which are equally weighted up to a phase.   However, 
when the connecting solution is Euclidean (or complex
with real Euclidean action), only the exponentially
{\it decreasing} solution contributes and the leading semiclassical
term is $e^{-|S_E|}$.  We follow the usual convention of $iS = -S_E$
so that this exponentially decreasing factor corresponds to the
stationary point with positive Euclidean action.

This seems to indicate a general property of the semiclassical
approximation (and perhaps of the entire expression) for \ref{mepi}
and \ref{mepiN} which we shall investigate further in section 
\ref{gen}.  Note that this case is more subtle than
that of the perturbed Bianchi I model
since that model had the
property that (for $m^2 >0$)
the Euclidean action of a Euclidean solution is always
positive when the `Euclidean lapse' $iN$ is positive.
In contrast, for 
the coupled harmonic oscillators, a Euclidean solution with
positive Euclidean lapse can have a Euclidean action with either sign.
This, therefore, is much closer to the generic case.

As in the perturbed Bianchi I model, we briefly compare \ref{home}
with a contour rotation scheme where $x^0$ is rotated to
$ix^0$.  Again, this gives just the right result 
in the `Euclidean sector' 
($a^2 > b^2$), but the branch cut of the square root creates an
ambiguity when analytically continuing back to the `Lorentzian sector'
$a^2 < b^2$.  Our result \ref{home} includes both possible
outcomes, equally weighted up to an $n$-dependent phase.

\section{The General Case}
\label{gen}

In sections \ref{frp} and \ref{cho}, we discoverd several interesting
properties of the exact results
\ref{spacelike}, \ref{timelike}, and \ref{home}.  We will now argue that
these properties should hold in general.  Specifically, we first 
show in section \ref{conv} 
that the inner product may be expressed as a path integral
over both `Lorentzian' {\it and} 
`Euclidean' paths (and combinations thereof) in which
the contribution of the `Euclidean' paths is
exponentially suppressed for small $\hbar$.  
Such an expression is more convergent
than \ref{mepi} and \ref{mepiN} and
may be of use for numerical investigations.  A representation of
this form also indicates that Lorentzian instantons (with either
sign of the action) and Euclidean instantons with positive Euclidean
action will contribute to our transition amplitude 
in the semiclassical limit.  We then argue in section
\ref{inst} that, in addition,
Euclidean instantons with negative action do {\it not}
contribute to the semiclassical limit.

\subsection{A simplified expression}
\label{conv}

Our general strategy will
be to simplify the path integral by performing
 the integrals in \ref{mepiN} over the momenta {\it
and} over the lapse; thus, we assume that we
may change the order of integration.  The result 
will (almost) be a configuration space path
integral in which, in effect, the constraint equation
\be
g_{ij}(x) {{\dot{x}^i \dot{x}^j} \over N^2} + V(x) = 0
\ee
has been solved for the lapse $N$.  That is, we will obtain
a path integral based on a `Baierlein-Sharp-Wheeler-like' \cite{BS} form
\be
\label{BSf}
S = \int dt \ \sqrt{g_{ij} \dot{x}^i \dot{x}^j V(x) }
\ee
of the classical action.

We begin with the expression \ref{mepiN} for the matrix elements
$\langle x | \delta(H) |x' \rangle$.  
Having written the integral in a form where Fadeev-Popov 
technology applies, we are now free to use any gauge fixing condition we
wish.
Because we would like to integrate out the momenta, we
will use a gauge condition of the form $G(x) = 0$ where
$G$ is a function of the coordinate variables {\it only} (see
\cite{HTV} for a discussion of canonical gauges in this context).  
In addition, we assume $\nabla G \cdot \nabla G >0$; that is, that
$\nabla G$ is `spacelike.'  
Unfortunately, $G=0$ is not really a `good' gauge
condition \cite{htgc} as there will always
exist points in phase space at which the Poisson bracket
$\{G,H\} = p \cdot \nabla G$ 
of $G$ with $H$ vanishes (such as where all momenta
vanish).  Nevertheless,
such a condition is often `good' on all but a set of measure zero.
Having admitted our treatment to be heuristic, we assume that
we may use such a gauge condition below.

Now, 
since there is only a single constraint, the Fadeev-Popov determinant takes
the particularly simple form of a product over times $t$:
\be
\Delta(\{G,H\}) = \prod_t |\{G,H\}(t)| = \prod_t |\nabla G(t) \cdot_t p(t)|
\ee
where the inner product $\cdot_t$ denotes a contraction through the
(co)metric $g^{ij}(x(t))$.  

For common factor orderings of the quantum constraint $H$, the
symbol $h(x,p)$ will be quadratic in the momenta, $p$.  We shall
assume that it takes the form
\be
h(x,p) = g^{ij}(x) p_{i} p_{j} + V_q(x)
\ee
where the subscript $q$ on $V$ indicates that the 
potential may receive `quantum corrections' such as terms proportional
to the curvature of $g_{ij}$ \cite{curve,curve2}.  In order to be explicit,
we shall assume the signature of $g$ to be everywhere 
$(1,n-1)$. 

Most of the the momentum integrals are 
of the form $\int_{-\infty}^{\infty} dp \ e^{i(p\dot{x} - \lambda p^2)}$ 
and can be performed exactly using stationary phase methods.  However, 
the presence of the absolute value $|\nabla G \cdot p|$ means that, 
for each value of $t$, there will be one momentum integral which
cannot be performed in this way.  It will be sufficient for our purposes
to leave this integral undone, and to simply perform the others.  
The component of
$p$ in the direction of $\nabla G(x(t))$ will be denoted $p_\parallel$.
Similarly, $n_{\parallel}(x(t))$ will be the 
unit vector in the direction of $\nabla G(x(t))$.
Performing these momentum integrals yields the expression:
\begin{eqnarray}
\langle x | \delta(H) | x' \rangle &=& \int DN Dx Dp_\parallel 
\exp \left[i{1\over 2} \int_0^1  dt  g^{\perp}_{ij} 
{{\dot{x}^i \dot{x}^j} \over N} - N(V_q(x)
+ p_\parallel^2) \right] \cr
& \times &
\prod_t e^{i{\pi \sign(N) \over
2} \left( {{n-3} \over 2} \right)} |N(t)|^{{1-n} \over 2)} 
\sqrt{-\det{g^\perp}} \
\delta [G(x(t)]  \ |\nabla G(x(t))| \ |p_\parallel |
\end{eqnarray}
where the usual product of normalization factors at each time has
been dropped (they may absorbed into the definition of the measure).
Here $g^{\perp}_{ij}$ is the induced metric on surfaces of constant
$G(x)$.

  Finally, we wish to perform the integrals over $N$.  As in the
explicitly soluble models discussed in section \ref{exact}, these
integrals may be expressed in terms of Bessel functions.  Note
that each such integral is of the form
\be
I = 
\int_{-\infty}^{\infty} dN  e^{i{{\pi}
\sign(N) \over
2} \left( {{n -3} \over 2} \right)} |N(t)|^{(1-n)/2}
\exp \left( i \left[ {1\over 2} g_{ij} {{\dot{x}^i \dot{x}^j} \over
N} - N ( V_q(x) + p_\parallel^2) \right] \right).
\ee
This is essentially the same integral with which we were faced in 
section \ref{frp}. Introducing 
$v \equiv V + n_\parallel^2 p_\parallel^2$,
$k \equiv g_{ij} \dot{x}^i \dot{x}^j$, and dropping 
constant normalization factors, the integral yields
\be
I = (k/v)^{(3-n)/4} K_{(n-3)/2}(\sqrt{kv})
\ee
when $kv >0$ and
\be
I = \pi \Re \left\{ (-k/v)^{(3-n)/4} e^{i \pi \left( {{n-2} \over 2}
\right)}  H^{(1)}_{(n-3)/2}(\sqrt{-kv})
\right\}
\ee
when $kv <0$.

As a result, the physical inner product may be written
\be
\label{mpip}
\langle x | \delta(H) | x' \rangle = 
\int Dx Dp_{\parallel} \prod_t \ I(k,v) \ \delta(G)
|\nabla G| | p_\parallel | \sqrt{-\det g^\perp}.
\ee
The connection to the Baierlein-Sharp-Wheeler-like \ref{BSf} should be clear:
since Bessel functions are exponentials at large arguments,
if the momenta
$p_\parallel$ were replaced by their semiclassical values
$\dot{x} \cdot \nabla G \sqrt{v/k}$ then
\ref{mpip} would be a path integral based on the action \ref{BSf}, although
with a rather complicated measure.

This expression may be thought of as a sum over both 
`Lorentzian' and `Euclidean' bits of path at each time.
Here, we say that a segment of path is Lorentzian when $kv >0$, so
that a real
lapse $N$ would be associated to this segment
by solving the constraint
\be
{k \over {N^2}} + v = 0.
\ee
Similarly, a bit of path with $k/v < 0$ is `Euclidean'
as the corresponding lapse is imaginary.  Note that this occurs
despite the original expression \ref{mepi}
being an integral only over
real classical lapse $N$.  As in our exactly solvable
models, the Lorentzian bits contribute with both signs of the 
action but the Euclidean path bits are always exponentially
{\it suppressed} when the corresponding $|S_E|$ is large.  
A similar result is obtained if other momentum 
integrals are left undone as well, so long as the momenta at each time
are integrated over at least one two-plane with signature $(-,+)$. 

The argument of every Bessel function is now manifestly
positive.  As a result, the fact that neither the Lorentzian nor
the Euclidean action is positive definite has disappeared
from sight.  If we define a slightly modified
function $I_\epsilon$ by $I_\epsilon(k,v) = I(k,v)$ for $kv>0$
and
\be
I_\epsilon(k,v) \equiv \Re \left[(-k/v)^{(3-n)/4} e^{i \pi \left(
{{n-2} \over 2} \right) }  
H^{(1)}_{(n-3)/2}([1 - i | \epsilon |] \sqrt{-kv}) \right]
\ee
for $kv < 0$, then $I_{\epsilon}(k,v)$ vanishes in the limit
of large $kv$ (for $\epsilon \neq 0$).  Our matrix elements may therefore be
expressed in the form
\begin{eqnarray} 
\label{cgt}
\langle x | \delta(H) | x' \rangle &=& \lim_{\epsilon \rightarrow 0}
\int Dx Dp_\parallel
\prod_t
I_{\epsilon}(k(t),v(t)) 
\cr & \times &
\delta(G(x(t)) \ |\nabla 
G(x(t))| \ |p_\parallel | \ \sqrt{-\det{g^\perp}}
\end{eqnarray}
for which the integrand is exponentially 
decreasing.  As it converges more strongly than \ref{mepi}, we may
hope that \ref{cgt} will be of use in numerical computations.

\subsection{Instantons and the Semiclassical Approximation}
\label{inst}

The expression \ref{mpip} for our transition amplitude includes
an explicit sum over both Lorentzian and Euclidean paths, as well
as arbitrary combinations of the two.
In the semiclassical approximation, this result
indicates that both Lorentzian and Euclidean stationary
points contribute to physical inner product, so long as the 
latter have positive Euclidean action.  In fact, since
the contribution of Euclidean path-bits is exponentially
suppressed in \ref{mpip} for small $\hbar$, it appears
that stationary points with 
negative Euclidean action {\it never} contribute,
as was suggested by the examples of section \ref{exact}.  However, we
have not yet shown this carefully.  In particular, it is possible
that a stationary point of the
action, while not actually on the contour over which the integral is
performed, might contribute to the semiclassical approximation\footnote{
We have already seen an example of such behavior as the Euclidean
paths that are explicitly included in \ref{cgt} correspond
to stationary points of the $N$ integrals at imaginary lapse (and
therefore off the original {\it real} contour).}. 
In general, in fact, we {\it do} expect such stationary points to
contribute, as the contour of integration could easily be deformed to
reach a complex stationary point lying just off the real axis.  If the
contour, say for the variable $k$ in \ref{mpip}, 
can be deformed far enough, it could wrap 
around the branch point at $k=0$ to reach a stationary
point on the `second sheet' of the Riemann surface on which $k$ lives.
Such a stationary
point would then contribute as $e^{+|S_E|}$ to 
the inner product $\langle x | \delta(H) | x' \rangle$.  We 
will now argue that this does not occur.  Because we study
the semiclassical approximation, the factors of $\hbar$ will be restored
below.
 
Let us first note that, if the semiclassical approximation is to hold, 
our matrix elements must be of the form
\be
\langle x |\delta(H)|x'\rangle = e^{iS(x,x')/\hbar}C(x,x')
\ee
where $C(x,x')$ is slowly varying in comparison with
$e^{iS(x,x')/\hbar}$.  In particular, since 
$S(x,x')$ is continuous on regions ${\cal R}$ of $\R^n \times \R^n$
where the number $n$ of connecting paths in constant, $C(x,x')$
must be continuous there as well.  

In general, we expect the union of the boundaries of such regions
${\cal R}$ to have measure zero so that 
the union $U$ of the interiors has measure
one.  We therefore {\it assume} that this is so and consider only
the open subset $U$ on which $\langle x | \delta(H) | x' \rangle$
is continuous.

If the matrix element $\langle x_0 | \delta(H) | x'_0 \rangle$
takes the form $e^{+|S_E(x_0,x_0')|/\hbar}$ (to leading semiclassical
order\footnote{The fact that a Euclidean path dominates does not
necessarily require the matrix elements to be real and positive;
there may in fact be an overall phase that we have neglected.  The point
is that this phase is both independent of $\hbar$ (for
small $\hbar$) and slowly
varying with $x$ and $x'$.}) at some pair $(x_0,x_0') \in U$, 
it must in fact take this form on some open rectangular
region $V_\hbar \times W_\hbar$ ($V_\hbar \subset \R^n$,
$W_\hbar \subset \R^n$) containing $(x_0,x_0')$.  If the diameters
of $V_\hbar,W_\hbar$ are much less than $\hbar \left[ \left| {{\partial
S_E} \over {\partial x}} \right| \right]^{-1}$ 
and $\hbar \left[ 
\left| {{\partial S_E} \over {\partial x'}} \right|  \right]^{-1}
$ respectively, then $S(x,x')/\hbar$
is essentially constant 
on $V_\hbar \times W_\hbar$.  As a result, $\langle x | \delta(x) |
x' \rangle $ is essentially real on $V_\hbar \times W_\hbar$ and
satisfies $\langle x | \delta(H) | x' \rangle \ge e^{\lambda/\hbar} $
where (say) $\lambda = {1 \over 2} |S_E(x,x')|$. 

We now choose two states
$|\phi \rangle$ and $|\psi \rangle$ in ${\cal S}$ whose representations
$\langle x | \phi \rangle = f({{x-x_0} \over \hbar})$, 
$\langle x | \psi \rangle = g({{x-x_0} \over \hbar})$ 
are positive
real, supported in $V_\hbar$ and $W_\hbar$ respectively, and have
$f(y)$ and $g(y)$ independent of $\hbar$.  Let us define
\begin{eqnarray}
& a_{\phi} =  \hbar^{-n} 
\int_{\R^n} \langle x | \phi \rangle d^nx   = \int_{\R^n} f(y) d^ny, & \cr
& a_{\psi} =  \hbar^{-n} 
\int_{\R^n} {\langle x | \psi \rangle}  d^nx  =
\int_{\R^n} g(y) d^ny  & \cr
& b_{\phi} = \hbar^{-n} 
\int_{\R^n} |\langle x | \phi \rangle|^2 d^nx  ,  \ 
b_{\psi} = \hbar^{-n} 
\int_{\R^n} |\langle x | \psi \rangle|^2 d^nx &.
\end{eqnarray}
It follows that $a_{\phi},a_{\psi},b_{\phi}, b_{\psi}$ are real, positive,
and independent of $\hbar$.  Note that the
physical inner product of our two states satisfies
\be
\label{grow}
\langle \phi_\hbar | \delta(H) | \psi_\hbar \rangle \ge
\hbar^{2n} a_{\phi}a_{\psi} e^{\lambda/\hbar}
\ee
to leading semiclassical order and that the expression on the right
diverges as $\hbar \rightarrow 0$. 

We will now derive a contradiction.  To do so, we return to our
original expression for the inner product
\be
\langle \phi_\hbar | \delta(H) | \psi_\hbar \rangle
= \int_{-\infty}^{+\infty} \langle \phi_\hbar | 
e^{-iHN/\hbar} | \psi_\hbar \rangle .
\ee
Recall that, for $|\phi_\hbar \rangle$, $|\psi_\hbar \rangle \in
{\cal S}$, this integral {\it converges} at large $N$.
Effectively, this is because the states $| \phi_\hbar \rangle$,
$|\psi_\hbar \rangle$ are characterized (to accuracy $\epsilon$) by
some minimal energy scale\footnote{This is the energy at which
$\Bigg| \langle \phi_\hbar | \delta(H)|\psi_\hbar \rangle  -
\langle \phi_\hbar | \delta(H - E_\hbar)|\psi_{\hbar} \rangle \Bigg| <
\epsilon$.  As a result, it is related to a sort of continuity
\cite{ban}
of the spectral representations of $|\phi_\hbar \rangle$ and 
$|\psi_\hbar \rangle$.  $E_\hbar$ depends on the functions $f$ and $g$
as well as on $\hbar$ and the accuracy $\epsilon$.  In particular, it
is not a property of the Hamiltonian $H$ alone and in no way
corresponds to a mass gap for the system.}
$E_\hbar$ and the integration over the region
$|N| > T_\hbar \approx \hbar/E_\hbar$ yields a
negligible contribution.  Note, however, that since
$|\phi_\hbar \rangle, |\psi_\hbar \rangle$ have norms 
$ \sqrt{\hbar^n b_\phi}$ and $ 
\sqrt{ \hbar^n b_\psi}$ and since the operator
$e^{-iHt/\hbar}$ is unitary, we have the bounds
\be
\langle \phi_\hbar | e^{-iHN/\hbar} | \psi_\hbar \rangle \leq \hbar^{n} 
\sqrt{b_\phi b_\psi}
\ee
and 
\be
\label{zerob}
\langle \phi_\hbar | \delta(H) |  \psi_\hbar \rangle \leq \hbar^{n+1} 
C \sqrt{b_\phi b_\psi} / E_\hbar
\ee
where $C$ is some constant independent of $\hbar$.
In order to compare this bound with the semiclassical expression
\ref{grow},  we estimate $E_\hbar$ as follows.

Recall that $h = g^{ij}(x) p_i p_j + V(x)$.  The states
$|\phi_\hbar \rangle$ and $|\psi_\hbar \rangle$ are characterized by 
coordinates $x_0$ and $x_0'$ (which are independent of $\hbar$)
and by momentum scales
\be
p_\hbar = {\hbar \over {{\rm diam} (V_\hbar)}} = \left[
{{\partial S_E} \over {\partial x}} \right], \ \ \ \ 
p'_\hbar = {\hbar \over {{\rm diam} (W_\hbar)}} = \left[
{{\partial S_E} \over {\partial x'}} \right] 
\ee
which are also independent of $\hbar$.  As a result, 
$E_\hbar \approx
g(x_0)p_\hbar^2 + V(x_0)$ is 
independent of $\hbar$ and the bound \ref{zerob}
{\it vanishes} as as $\hbar \rightarrow 0$, in contradiction with
\ref{grow}.  Here, $g(x_0)$ is some bound on the components of 
the $g^{ij}(x)$ at $x =x_0$.
We thus conclude that, at least for minisuperspace models, Euclidean
instantons always contribute to our transition amplitude 
as $e^{-|S_E|/\hbar}$.  Similarly, complex instantons must contribute
as $e^{i \Re S/\hbar} e^{- | \Im S/\hbar | }$, where $\Im$ denotes the
imaginary part.

\section{Discussion}
\label{disc}

The goal of this work was to use a canonical formalism (the
refined algebraic or Rieffel induction method) to 
represent a minisuperspace transition amplitude as a path integral and to 
investigate this expression both exactly and in the semiclassical limit.
We have seen that the path integral may be written in the 
form \ref{cgt} which explicitly sums over both `Lorentzian path-bits'
(which appear with both signs of the action) and `Euclidean path-bits'
which are exponentially suppressed.  As a result, we conclude
that any Lorentzian or positive action Euclidean stationary
point may contribute to our transition amplitude in the semiclassical
limit.  We also argued in section \ref{inst} that Euclidean
instantons with {\it negative} Euclidean action (which would be
exponentially enhanced) do {\it not}
contribute in the semiclassical limit.

It seems reasonable to assume that similar
results hold in full $3+1$ gravity, Kaluza-Klein theory, 
dilaton gravity, or any other diffeomorphism invariant theory
of gravity with indefinite Euclidean action.  However, it should be noted
that the most interesting instantons in quantum gravity (e.g., 
\cite{pair1,pair2,pair3,vac,nb1,nb2,HH,JJ1,JJ3,Baum,Hawk,Cole})
are rather far from our minisuperspace models.  In particular, they
involve processes in which the spatial topology of the initial
state differs from the spatial topology of the final state;
an effect which appears to be ruled out by construction in our
context (see however, \cite{JJ1,JJ3}).  An argument
could certainly be made that such instantons are qualitatively
different and that our results have no bearing on their interpretation.
On the other hand, we note that the arguments of \ref{inst}
did not rely on the details of our model, but followed from the
unitarity
of $e^{-iHN}$ and the existence of an appropriate subspace ${\cal S}
\subset \haux$.  One would expect such arguments to generalize readily
if refined algebraic/Rieffel induction methods are applicable at all.
At the very least, our
arguments are suggestive, and it is worthwhile to briefly discuss below
the hypothesis that our results do generalize to such cases.  

Many instanton calculations, such as the pair creation calculations
of \cite{pair1,pair2,pair3}, 
implicitly assume that the relevant stationary point is
the one with positive Euclidean {\it lapse}.  So long as the 
corresponding Euclidean
action is positive (as it is in all of the pair creation
examples \cite{pair1,pair2,pair3}), 
this is just the conclusion that would follow from section \ref{inst}.
We will therefore concentrate on situations where the Euclidean action
is {\it negative}.  Perhaps the most interesting use of such instantons
was seen in the arguments of Baum \cite{Baum}, Hawking\cite{Hawk}, and 
Coleman \cite{Cole}
that the cosmological constant $\Lambda$ should vanish.
They supposed that, in some way, the quantum state of the universe
provides a probability distribution for $\Lambda$ and
proceeded to estimate this distribution through instanton calculations.
The large negative action of four-sphere instantons for small
positive $\Lambda$ was used to argue that this distribution is
exponentially large (or even larger \cite{Cole}) for
$\Lambda$ near zero.

While we have been
interested in the transition amplitude and not in a particular quantum
state, we note that {\it every} physical state
$|\psi_{phys} \rangle$ in our scheme may be expressed in the form
\be
\langle x | \psi_{phys} \rangle = \int dx' \langle x| \delta(H) | x'
\rangle \langle x | \psi \rangle
\ee
for some $|\psi \rangle \in {\cal S}$.  As a result, the arguments of
section \ref{inst} imply that no physical state has a
`wavefunction' $\langle x | \psi_{phys} \rangle$ which behaves as
$e^{|S_E(x)|}$ in any region of superspace.  This suggests
that the instantons of \cite{Baum,Hawk,Cole} would not
in fact contribute to the desired distribution.
Following a similar line of reasoning, \cite{steve,steve1} arrives at
this same conclusion.  Other arguments against and comments
on the Baum-Coleman-Hawking mechanism include 
\cite{anti1,anti2,anti3,anti4,anti5}.

Another point to be addressed is the argument of
\cite{HH} that only stationary points with $\Re \sqrt{g_E} \ge 0$
(where $g_E$ is the Euclidean metric) should contribute to the path
integral.  In our notation this restriction would be $\Re(iN) \ge 0$; i.e.,
positive Euclidean lapse.  The
argument of \cite{HH}
was based on considering the normalizability of an induced
quantum state for the matter (non-gravitational) fields.  The issue
concerned whether this state is an exponentially growing or decaying 
function of these fields; that is, it had
to do with the state functional at large values of the matter fields.
While we have argued that 
stationary points with $\Re(iN) <0$ may contribute, the fact that
the Euclidean action of matter fields is positive for $\Re(iN)>0$
means that for large values of the matter fields our condition of
positive Euclidean action is equivalent to the condition 
$\Re(iN) >0$ of \cite{HH}.  This is as one would expect since 
exponentially increasing functions do not define generalized eigenstates
of the constraint.  As a result, we see that
our arguments are consistent
with the general requirements of \cite{HH} for reproducing 
quantum field theory in curved spacetime.

\acknowledgments
This project sprouted from a conversation with Andrew Chamblin, to
whom the author would like to express his gratitude.  The author would also
like to thank Andrei Barvinsky, Jim Hartle, Gary Horowitz, 
Joe Polchinski, and especially Jorma Louko for helpful
comments and discussions.   This work was supported by NSF grant
PHY95-97965.

\appendix

\section{On the Convergence of the Path Integral Representations}
\label{cds}

In this appendix we discuss in detail the convergence properties of 
the path integrals \ref{mepi} and \ref{mepiN}
for the physical inner product.   These expressions do in fact
converge, so long as all integrations are appropriately interpreted.

Let us begin by noting that $\langle \phi | e^{-iHN}|\psi \rangle$ is
well-defined for all $|\phi \rangle$, $ |\psi \rangle \in \haux$, and
in particular for all $|\phi \rangle$, $|\psi \rangle \in {\cal S}$.
This follows from the fact that $e^{iHN}$ is a unitary (and therefore
bounded) operator on $\haux$.
As a result, $\langle x|e^{-iHN}| x' \rangle$ is a well-defined
distribution on ${\cal S} \times {\cal S}$; that is, 
$\langle x | e^{-iHN} |x' \rangle$ is a member of 
the dual space ${\cal S}' \times {\cal S}'$.

Note that this holds for {\it all} real $N$, even though the expression
for $\langle x|e^{-iHN}|x'\rangle$ (for the case $H= -P^2_0 + P^2_1 =
m^2$, for example) as a function of $N$, $x$, and $x'$ may have
an essential singularity at $N=0$ (see, e.g. \cite{JJ1,JJ3}).  
This is nothing more than the
fact that the distribution $\delta(x-x')$ cannot be represented
by a smooth function.  In fact, $\langle x| e^{-iHN} | x' \rangle$
is a {\it continuous} function of $N$ in the topology of ${\cal S}' \times 
{\cal S}'$, even at $N=0$.  There is thus no difficulty with the
$N$ integral at any finite value 
of $N$.

We expect the result of this integral to be $\langle x |\delta(H) |
x' \rangle$.  Our
technical assumptions guarantee 
that the limit
\be
\lim_{{a \rightarrow -\infty} \atop {b \rightarrow +\infty}} \int_a^b dN
\langle x|e^{-iHN}| x'\rangle
\ee
does in fact converge to $\langle x |\delta(H)|x'\rangle$ (in the topology of
${\cal S}' \times {\cal S}'$), and does so independently of
how the limits of $a$ and $b$
are taken \cite{ban}.  Any lack of 
convergence
of \ref{mepi} must therefore
arise from the path integral representation of
$\langle x | e^{-iHN} | x' \rangle$.  It is to this expression that we
now turn.

Our assumptions (from \cite{ban}) guarantee that $(1 - iHt)$ maps 
${\cal S}$ into ${\cal S}$ so that the operator $(1 - iH N/k)^k$ is 
also a member of ${\cal S}' \times {\cal S}'$.  It follows that
the integrations
in the k-skeletonized path integral
\be
I_k = \int {{dp_0} \over {2 \pi}}
\prod_{i=1}^k {{dp(t_i) dq(t_i)} \over {2 \pi}}
e^{i p(t_i) [q(t_i) - q(t_{i-1})]}
\left( 1 - i {N \over k} h(p(t_i),q(t_i))\right)
\ee
must converge in the sense of defining elements of ${\cal S}' \times 
{\cal S}'$.

In addition, for fixed states $|\phi \rangle \in {\cal S} $ and 
$|\psi \rangle \in {\cal S}$, we have
\be
\langle \phi | e^{-iHt} |\psi \rangle - \langle \phi | (1-iHt)| \psi \rangle
\le C t^2
\ee
for small $t$ and
some (state-dependent) constant $C$.  Thus, the 
$k \rightarrow \infty$ limit of the k-skeletonized path integral
converges (as a sequence in ${\cal S}' \times {\cal S}'$) to 
$\langle x | e^{-iHN} | x' \rangle$.  We conclude that, when 
appropriately interpreted, the expression \ref{mepi} 
does in fact converge, despite the oscillatory nature of the exponential
and the fact that the `Euclidean action' is unbounded below.  
Since \ref{mepiN} is designed to exactly reproduce
\ref{mepi}, we have established the convergence of this expression as
well.

\section{Stationary Points and Analyticity of the Measure} 
\label{sc}

We now show how 
the the analytic structure of the measure in \ref{frpint}
determines the overall form
of the results \ref{spacelike} and \ref{timelike}
by evaluating this result in the
semiclassical approximation.  The structure of the results
\ref{home} and \ref{mpip} is determined in essentially the same way.

Consider the integral \ref{frpint}:
\begin{eqnarray}
\label{sci}
\langle x | \delta(H) | x' \rangle  &=& {1 \over {2 \pi} }
\int_{-\infty}^{\infty} dN {{e^{i {\pi \over 2} (1 - {n \over 2}) {\rm
sign}(N)}} \over {(2\pi |N|)^{n/2}}} \exp \left(
-i{{m^2} \over 2}  \left[ N - {{ (x - x')^2} \over {m^2 N}} \right]
\right)  \cr
& = & {1 \over {2 \pi}} \int_{-\infty}^{\infty} dN \ 
{{e^{-i \pi n/4}} \over {(2 \pi N)^{n/2}}} e^{(i \pi /2) {\rm sign} (N)}
\exp \left( -i{{m^2} \over 2}  \left[ N - {{ (x - x')^2} 
\over {m^2 N}} \right] \right)
\end{eqnarray}
where the function $N^{n/2}$ is $(\sqrt{N})^n$ and $\sqrt{N}$ is defined
to have a branch cut along the {\it positive} imaginary axis.
Note that ${\rm sign}(N)$ may be interpreted as an
analytic function which is constant on the left and right half
planes with a cut along the entire imaginary axis.

Now, for large $N$ (and $m^2 >0$), the exponential factor decays
in the lower half plane but grows in the upper half plane.  As 
a result, the contour may be closed in the lower half plane as shown
below.

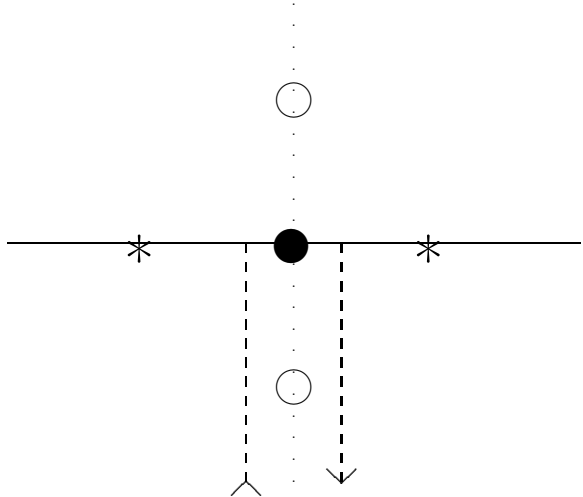
\setlength{\unitlength}{0.01250000in}%
\begingroup\makeatletter\ifx\SetFigFont\undefined
\def\x#1#2#3#4#5#6#7\relax{\def\x{#1#2#3#4#5#6}}%
\expandafter\x\fmtname xxxxxx\relax \def\y{splain}%
\ifx\x\y   
\gdef\SetFigFont#1#2#3{%
  \ifnum #1<17\tiny\else \ifnum #1<20\small\else
  \ifnum #1<24\normalsize\else \ifnum #1<29\large\else
  \ifnum #1<34\Large\else \ifnum #1<41\LARGE\else
     \huge\fi\fi\fi\fi\fi\fi
  \csname #3\endcsname}%
\else
\gdef\SetFigFont#1#2#3{\begingroup
  \count@#1\relax \ifnum 25<\count@\count@25\fi
  \def\x{\endgroup\@setsize\SetFigFont{#2pt}}%
  \expandafter\x
    \csname \romannumeral\the\count@ pt\expandafter\endcsname
    \csname @\romannumeral\the\count@ pt\endcsname
  \csname #3\endcsname}%
\fi
\fi\endgroup
\centerline{
\begin{picture}(244,210)(138,452)
\thinlines
\multiput(274,465)(0.40000,-0.40000){16}{\makebox(0.1111,0.7778){\SetFigFont{5}{6}{rm}.}}
\multiput(280,459)(0.40000,0.40000){16}{\makebox(0.1111,0.7778){\SetFigFont{5}{6}{rm}.}}
\multiput(234,454)(0.40000,0.40000){16}{\makebox(0.1111,0.7778){\SetFigFont{5}{6}{rm}.}}
\multiput(240,460)(0.40000,-0.40000){16}{\makebox(0.1111,0.7778){\SetFigFont{5}{6}{rm}.}}
\put(260,620){\circle{14}}
\put(260,500){\circle{14}}
\put(259,559){\circle*{14}}
\multiput(240,460)(0.00000,8.00000){13}{\line( 0, 1){  4.000}}
\multiput(240,560)(7.27273,0.00000){6}{\line( 1, 0){  3.636}}
\multiput(280,560)(0.00000,-8.00000){13}{\line( 0,-1){  4.000}}
\multiput(260,460)(0.00000,9.09091){23}{\makebox(0.1111,0.7778){\SetFigFont{5}{6}{rm}.}}
\put(140,560){\line( 1, 0){240}}
\put(189,543){\makebox(0,0)[lb]{\smash{\SetFigFont{34}{40.8}{rm}*}}}
\put(310,543){\makebox(0,0)[lb]{\smash{\SetFigFont{34}{40.8}{rm}*}}}
\end{picture}
}
On this diagram, the filled black circle
represents the essential singularity at $N=0$,
the open circles are the Euclidean stationary points at
$N = \pm \sqrt{{-(x-x')^2 } \over {m^2}}$ when $(x-x')^2 >0$, and 
the $*$'s are the Lorentzian
stationary points at $N = \pm \sqrt{ {  -(x - x')^2} \over {m^2}}$
for $(x-x')^2 < 0$.  The vertical dotted line is the cut 
that defines ${\rm sign}(N)$, while the horizontal line represents
the real axis -- the original contour of integration.  The dashed
line going up and then back down the negative imaginary axis is a
new contour to which the original one may be deformed.
Note that, because of the cut,
the integrations along the negative imaginary axis do not cancel, but
instead give equal contributions and add together.  Since, for
$(x - x')^2 >0$, the integrand vanishes as $N$ approaches zero from the
lower half plane, the essential singularity at $N=0$ does not
contribute in the semiclassical limit.  This limit is therefore 
dominated by the stationary point in the lower half-plane, for which the
integrand in eq. \ref{sci} is exponentially suppressed by
a factor of $e^{-m \sqrt{(x-x')^2}}$ in agreement
with the leading behavior of \ref{spacelike}.

\end{document}